\documentclass[12pt,a4paper]{iopart} 
\usepackage{iopams}
\usepackage{setstack,cite}
\usepackage{amssymb,graphicx}

\newcommand{\bea}{\begin{eqnarray}}
\newcommand{\eea}{\end{eqnarray}}
\newcommand{\ds}{\displaystyle}
\newcommand{\al}{\alpha}

\newcommand{\del}{\delta}

\begin{document}
\title[Bright solitons in multicomponent LSRI system]
{Higher dimensional bright solitons and their collisions in multicomponent long wave-short wave system} 
\author{T. Kanna$^1$, M. Vijayajayanthi$^2$, K. Sakkaravarthi$^1$ and M. Lakshmanan$^2$}
\address{$^1$ Department of Physics,  Bishop Heber College,\\Tiruchirapalli--620 017, India\\}
\address{
$^2$ Centre for Nonlinear Dynamics, School of Physics, Bharathidasan University, Tiruchirapalli--620 024, India}
\ead{lakshman@cnld.bdu.ac.in}
\begin{abstract}
Bright plane soliton solutions of an integrable (2+1) dimensional ($n+1$)-wave system are obtained by applying Hirota's bilinearization method.  First, the soliton solutions of a 3-wave system consisting of two short wave components and one long wave component are found and then the results are generalized to the corresponding integrable ($n+1$)-wave system with $n$ short waves and single long wave.  It is shown that the solitons in the short wave components (say $S^{(1)}$ and $S^{(2)}$) can be amplified by merely reducing the pulse width of the long wave component (say L).  The study on the collision dynamics reveals the interesting behaviour that the solitons which split up in the short wave components undergo shape changing collisions with intensity redistribution and amplitude-dependent phase shifts.  Even though similar type of collision is possible in (1+1) dimensional multicomponent integrable systems, to our knowledge for the first time we report this kind of collisions in (2+1) dimensions.  However, solitons which appear in the long wave component exhibit only elastic collision though they undergo amplitude-dependent phase shifts.
\end{abstract}
\maketitle
    
\section{Introduction}
One of the main emphasis of current research in the area of integrable systems and their 
 applications is the study on multicomponent nonlinear systems admitting soliton type solutions \cite{ref1,ref2,ref3,ref3a,ref4,ref5,ref6,ref16,ref6a,ref7,ref8,ref10,ref10a,ref11,ref12,ref13,ref13a}.  In (1+1) dimensions, it has been shown that the multicomponent bright solitons of the integrable N-coupled nonlinear Schr\"odinger (CNLS) equations undergo fascinating shape 
 changing collisions with intensity redistribution which have no single component 
 counterpart \cite{ref3a,ref4,ref5,ref6,ref16}.  This interesting behaviour found applications in nonlinear switching 
 devices\cite{ref10}, matter wave switches \cite{ref10a} and more importantly in the context of optical computing in bulk 
 media \cite{ref11,ref12}.  There is a natural tendency to look for such kind of collisions in higher 
 dimensions.  From this point of view, we have considered the following recently studied integrable coupled  (2+1) dimensional ((2+1)D) system, which is a two component 
 analogue of the two dimensional long wave-short wave resonance interaction (LSRI) system \cite{ref13}, in dimensionless form, 
\numparts  
\bea
&&i(S_t^{(j)}+S_y^{(j)})-S_{xx}^{(j)}+L S^{(j)}=0, \quad j=1,2,\\
&&L_t=2\sum_{j=1}^2 |S^{(j)}|^2_x,
\eea
\endnumparts
where the subscripts denote partial derivatives.  (Note that in the above two components mean two short wave ($S$) components).  The one component ($j=1$) version of the above equations corresponds to the interaction of a long interfacial wave (L) and a short surface wave (S) in a two layer fluid \cite{ref14}.  Also in ref. \cite{radha}, the existence of dromion like solution was established for the $j=1$ case.  In their very recent interesting work, Ohta, Maruno and Oikawa \cite{ref13} have derived equation (1) as the governing equations for the interaction of three nonlinear dispersive waves by applying a reductive perturbation method.  Here, among these three waves, two waves are propagating in the anomalous dispersion region 
and the third wave is propagating in the normal dispersion regime.  In 
the context of long wave-short wave interaction, the first two components can be viewed 
as the two components of the short surface waves while the last component corresponds to the long 
interfacial wave.  Note that the presence of the long interfacial wave induces the nonlinear interaction between the two short wave components which leads to nontrivial collision behaviour as will be shown in this paper.  Here onwards we call equation (1) as the 3-wave LSRI system in which the first two components
 correspond to short waves and the last one is a long wave.

Apart from deriving the governing equation (1) in ref. \cite{ref13}, Ohta \etal have also given Wronskian type soliton solutions of a specific type where the components $S^{(1)}$,  $S^{(2)}$, and $L$ comprise of $N$ solitons, $M$ solitons, and $(M+N)$  solitons, respectively.  In this context, however, it is of considerable interest to study the 
collision behaviour if the same number of solitons are split up in all the three components and to check whether nontrivial shape changing collisions of solitons as in the case of CNLS systems \cite{ref3a,ref4} occur here also and to look for the possibilities of construction of logic gates based on soliton collisions.  For 
the one component case the interaction of two solitons in both short wave and long wave 
components have been studied in detail in ref. \cite{ref14} and certain interesting features such as fusion and fission processes have been 
revealed.  In this study, we consider the multicomponent (2+1)D LSRI system admitting the same number of bright solitons in all the three components and obtain the multisoliton solutions.  Our analysis on their collision properties shows that the solitons appearing in the short wave components exhibit a shape changing collision scenario resulting in a redistribution of intensity as well as amplitude-dependent phase shift whereas the long wave component solitons undergo standard elastic collisions only but with amplitude-dependent phase shifts.  We also point out that the $(N, M, N+M)$ soliton solutions obtained in ref. \cite{ref13} follow as special cases of the $(m, m, m)$ multisoliton solution obtained here when some of the soliton parameters are restricted to very special values.  The study is also extended to the ($n+1$) wave system as well, where $n$ is arbitrary. 

The plan of the paper is as follows:  In section 2, we briefly present the bilinearization procedure for the three wave system.  Multisoliton solution of the three wave system is discussed in section 3.  Explicit one-soliton and two-soliton solutions are analysed in section 4. The asymptotic analysis of the two-soliton solution of the three wave system is given in section 5.  The interesting collision scenario of two solitons is discussed in detail in section 6.  Sections 7 and 8 deal with three- and four-soliton solutions, respectively.  Multicomponent case with $j>2$ in equation (1) is studied in section 9.  Section 10 is allotted for conclusion. 

\section{(2+1)D bright soliton solutions}
The soliton solutions of equation (1) are obtained by using Hirota's direct method \cite{ref15,hirotabook}.  By performing the bilinearizing transformations
\bea
S^{(j)}=\frac{g^{(j)}}{f}, \quad L=-2\frac{\partial^2}{\partial x^2}(\log{f}), \quad j=1,2,
\eea
where $g^{(j)}$'s are complex functions while $f$ is a real function, 
equation (1) can be decoupled into the following bilinear equations  
\numparts
\bea
&&\left(i(D_t+D_y)-D_x^2\right)(g^{(j)}\cdot f)=0,\quad j=1,2,\\
&&D_tD_x(f \cdot f)=-2\sum_{\substack{j=1}}^{2} (g^{(j)}g^{(j)*}), 
\eea
\endnumparts
where $*$ denotes the complex conjugate.  The Hirota's bilinear operators $D_x$, $D_y$ and $D_t$ are defined as 
\bea
\hspace{-2.5cm}D_x^{p}D_y^{q}D_t^{r}(a\cdot b) =\bigg(\frac{\partial}{\partial x}-\frac{\partial}{\partial x'}\bigg)^p\bigg(\frac{\partial}{\partial y}-\frac{\partial}{\partial y'}\bigg)^q \bigg(\frac{\partial}{\partial t}-\frac{\partial}{\partial t'}\bigg)^r a(x,y,t)b(x',y',t')\Big|_{ (x=x',y=y', t=t')}.\nonumber
\eea  
Expanding $g^{(j)}$'s and $f$ formally as power series expansions in terms of a small arbitrary real parameter $\chi$,
\numparts
\bea
g^{(j)}&=&\chi g_1^{(j)}+\chi^3 g_3^{(j)}+\ldots, \quad j=1,2,\\
f&=&1+\chi^2 f_2+\chi^4 f_4+\ldots,
\eea
\label{ps}
\endnumparts
and solving the resultant set of linear partial differential equations recursively, one can obtain the explicit forms of $g^{(j)}$ and $f$.  Then by substituting their expressions in (2) one can write down the soliton solutions.  The procedure has been successfully used to unearth several interesting properties of soliton collisions associated with CNLS system in Refs. \cite{ref3a,ref4,ref5,ref6,ref16}.  We have used a similar procedure here and obtained the one-soliton (1, 1, 1), two-soliton (2, 2, 2), three-soliton (3, 3, 3) and four-soliton (4, 4, 4) solutions explicitly.  This can be generalized to the arbitrary $m$-soliton $(m, m, m)$ solution, in a Gram determinant form.  From this one may claim that in the general case the number of solitons which split up in the short wave components ($S^{(1)}$ and $S^{(2)}$) as well as in the long wave component (L) are the same.  However, we also point out that the (1, 1, 2), (2, 2, 4) and $(N, M, N+M)$ soliton solutions obtained by Ohta \etal \cite{ref13} can be deduced as special cases of our $(m, m, m)$ soliton solution with $m=2, m=4$ and $m=N+M$, respectively, for particular choices of parameters in the solutions.

\section{Arbitrary m-soliton solution}
We first present the general form of $(m, m, m)$ soliton solution for arbitrary $m$ in the following Gram determinant form.  In order to write down the multisoliton ($m$-soliton) solution of the three wave LSRI system (1), we define the following $(1\times m)$ row matrix $C_s$, $s=1,2$, $(m\times 1)$ column matrices $\psi_j$, and $\phi$, $j=1,2,\ldots, m$, and the $(m\times m)$ identity matrix $I$:
\numparts
\bea
C_s= -\left(\alpha_1^{(s)}, \alpha_2^{(s)}, \ldots, \alpha_{m}^{(s)}\right),\quad  {\bf{0}}=(0, 0, \ldots, 0),
\eea
\bea
\psi_j=\left(
\begin{array}{c}
\alpha_j^{(1)}\\
\alpha_j^{(2)}
\end{array}
\right), \quad \phi=\left(
\begin{array}{c}
e^{\eta_1}\\
e^{\eta_2}\\
\vdots\\
e^{\eta_{m}}\\
\end{array}
\right),\quad I=
\left(
\begin{array}{cccc}
1 & 0 & \cdots & 0\\
0 & 1 & \cdots & 0\\
\vdots & \vdots & \ddots &\vdots\\
0 & 0 & \cdots &1
\end{array}
\right).
\eea
\endnumparts
Here $\alpha_{j}^{(s)}$, $s=1,2$, $j=1,2,\ldots, m$, are arbitrary complex parameters and $\eta_i=k_ix-(ik_i^2+\omega_i)y+\omega_it$, $i=1,2, \ldots, m$, and $k_i$ and $\omega_i$ are complex parameters.  Then we can write down the multisoliton solution of the three wave LSRI system in the form of equation (2), with  
\bea
S^{(s)}=\frac{g^{(s)}}{f}, \quad s=1, 2, \quad L=-2\frac{\partial^2}{\partial x^2}\log(f), 
\eea
where
\numparts
\bea
g^{(s)}=
\left|
\begin{array}{ccc}
A & I & \phi\\
-I & B & {\bf 0}^T\\
{\bf 0} & C_s & 0
\end{array}
\right|, \quad \quad f= \left|
\begin{array}{cc}
A & I\\
-I & B 
\end{array}
\right|,
\eea
in which $s$ denotes the short wave components.  Here the matrices $A$ and $B$ are defined as 
\bea
\hspace{-2.6cm}A_{ij}= \frac{e^{\eta_i+\eta_j}}{k_i+k_j^*},\;B_{ij}&&=\kappa_{ji}=\frac{-\psi_i^{\dagger} \psi_j}{(\omega_i^*+\omega_j)}
=-\frac{(\alpha_j^{(1)} \alpha_i^{(1)^*}+\alpha_j^{(2)} \alpha_i^{(2)^*})}{(\omega_i^*+\omega_j)},\; i,j=1, 2, \ldots, m.
\label{omg}
\eea
\endnumparts
In equation (\ref{omg}),  $\dagger$ represents the transpose conjugate and the real parts of $\omega_i$'s (or $k_i$'s) should be chosen as negative quantities in order to obtain nonsingular solutions, which are necessary conditions.  Sufficiency condition requires the choice of parameters such that $f$ is real and nonzero (see below sections 4, 7 and 8 for details in the case of $m=1, 2, 3,$ and 4).
\subsection{Proof of multisoliton solution of the three wave LSRI system}
We now prove that the Gram determinant forms of $g^{(s)}$ and $f$ given above indeed satisfy the bilinear equations (3).  By applying the derivative formula for the determinants, that is,
\numparts
\bea
\frac{\partial D}{\partial x}=\sum_{1\leq i, j\leq n}\frac{\partial a_{i, j}}{\partial x}\frac{\partial D}{\partial a_{i, j}}=\sum_{1\leq i, j\leq n}\frac{\partial a_{i, j}}{\partial x} \Delta_{i, j},
\eea
where $D=
\left|
\begin{array}{cccc}
a_{11} & a_{12} & \cdots & a_{1n}\\
a_{21} & a_{22} & \cdots & a_{2n}\\
\vdots & \vdots & \ddots &\vdots\\
a_{n1} & a_{n2} & \cdots &a_{nn}
\end{array}\right|$ and $\Delta_{i, j}$ is the cofactor of the $(i, j)^{\mbox{th}}$ element and making use of the properties of bordered determinants and also the elementary properties of determinants \cite{hirotabook,proof}, the derivatives $g^{(s)}_x$, $f_x$, $f_t$, $f_{xt}$, $g^{(s)}_z$, $g^{(s)}_{xx}$, $f_z$ and $f_{xx}$, where $\frac{\partial}{\partial z}=\left(\frac{\partial}{\partial t}+\frac{\partial}{\partial y}\right)$, can be derived as below:

\bea
g^{(s)}_{x}=
\left|
\begin{array}{cccc}
A & I & \phi & \phi_x \\
-I & B & {\bf 0}^T & {\bf 0}^T\\
{\bf 0} & C_s & 0&0\\
{\bf 0} & {\bf 0} & -1 & 0
\end{array}
\right|, \quad \quad f_x= \left|
\begin{array}{ccc}
A & I & \phi \\
-I & B & {\bf 0}^T\\
-\phi^\dagger & {\bf 0} & 0
\end{array}
\right|,
\eea

\bea
\hspace{-1.5cm}f_{t}=-\sum_{s=1}^2
\left|
\begin{array}{ccc}
A & I & {\bf 0}^T \\
-I & B & -C_s^{\dagger}\\
{\bf 0} & C_s & 0 
\end{array}
\right|, \quad \quad 
f_{xt}=-\sum_{s=1}^2
\left|
\begin{array}{cccc}
A & I & \phi &{\bf 0}^T \\
-I & B & {\bf 0}^T& -C_s^{\dagger}\\
-\phi^{\dagger}&{\bf 0} & 0 & 0 \\
{\bf 0}&C_s&0 &0
\end{array}
\right|,
\eea

\bea
g^{(s)}_z=-i
\left|
\begin{array}{cccc}
A & I & \phi &\phi_{xx}\\
-I & B & {\bf 0}^T& {\bf 0}^T\\
{\bf 0} & C_s & 0 &0\\
{\bf 0}& {\bf 0}&-1 &0
\end{array}
\right|+i
\left|
\begin{array}{cccc}
A & I & \phi &\phi_{x}\\
-I & B & {\bf 0}^T& {\bf 0}^T\\
{\bf 0} & C_s & 0 &0\\
\phi^{\dagger}& {\bf 0}&0 &0
\end{array}
\right|
, 
\eea
\bea
g^{(s)}_{xx}=
\left|
\begin{array}{cccc}
A & I & \phi &\phi_{xx}\\
-I & B & {\bf 0}^T& {\bf 0}^T\\
{\bf 0} & C_s & 0 &0\\
{\bf 0}& {\bf 0}&-1 &0
\end{array}
\right|+
\left|
\begin{array}{cccc}
A & I & \phi &\phi_{x}\\
-I & B & {\bf 0}^T& {\bf 0}^T\\
{\bf 0} & C_s & 0 &0\\
\phi^{\dagger}& {\bf 0}&0 &0
\end{array}
\right|
, 
\eea
\bea
f_{z}=-i
\left|
\begin{array}{ccc}
A & I & \phi_x \\
-I & B & {\bf 0}^T\\
-\phi^{\dagger} & {\bf 0} & 0 
\end{array}
\right|+i
\left|
\begin{array}{ccc}
A & I & \phi \\
-I & B & {\bf 0}^T\\
-\phi^{\dagger}_x & {\bf 0} & 0 
\end{array}
\right|,
\eea
and
\bea
f_{xx}=
\left|
\begin{array}{ccc}
A & I & \phi_x \\
-I & B & {\bf 0}^T\\
-\phi^{\dagger} & {\bf 0} & 0 
\end{array}
\right|+
\left|
\begin{array}{ccc}
A & I & \phi \\
-I & B & {\bf 0}^T\\
-\phi^{\dagger}_x & {\bf 0} & 0 
\end{array}
\right|.
\eea
The conjugate of $g^{(s)}$ can be written as  
\bea
g^{(s)*}=
-\left|
\begin{array}{ccc}
A & I & {\bf 0}^T\\
-I & B & -C_s^{\dagger}\\
-\phi^\dagger & {\bf 0} & 0
\end{array}
\right|.
\eea
Substituting for $g^{(s)}_x$, $g^{(s)}_z$, $g^{(s)}_{xx}$, $f_x$, $f_{xx}$, and $f_z$ in equation (3a), we find
\bea
\left|
\begin{array}{cccc}
A & I & \phi &\phi_{x}\\
-I & B & {\bf 0}^T& {\bf 0}^T\\
{\bf 0} & C_s & 0 &0\\
-\phi^{\dagger}& {\bf 0}&0 &0
\end{array}
\right|
\left|
\begin{array}{cc}
A & I\\
-I & B 
\end{array}
\right|
=
\left|
\begin{array}{ccc}
A & I & \phi_x \\
-I & B & {\bf 0}^T\\
-\phi^{\dagger} & {\bf 0} & 0 
\end{array}
\right|
\left|
\begin{array}{ccc}
A & I & \phi\\
-I & B & {\bf 0}^T\\
{\bf 0} & C_s & 0
\end{array}
\right|\nonumber\\&&\hspace{-6cm}
-
\left|
\begin{array}{ccc}
A & I &  \phi_x \\
-I & B  & {\bf 0}^T\\
{\bf 0} & C_s & 0
\end{array}
\right|
 \left|
\begin{array}{ccc}
A & I & \phi \\
-I & B & {\bf 0}^T\\
-\phi^\dagger & {\bf 0} & 0
\end{array}
\right|.
\eea 
This is nothing but a Jacobian identity and hence $g^{(s)}$ and $f$ satisfy the first bilinear equation (3a).  In a similar way one can also check that the second bilinear equation (3b) gives rise to the following Jacobian identity 
for the Gram determinant forms of $g^{(s)}$ and $f$:
\bea
\hspace{-2cm}
-\sum_{s=1}^2
\left|
\begin{array}{cccc}
A & I & \phi &{\bf 0}^T \\
-I & B & {\bf 0}^T& -C_s^{\dagger}\\
-\phi^{\dagger}&{\bf 0} & 0 & 0 \\
{\bf 0}&C_s&0 &0
\end{array}
\right|
\left|
\begin{array}{cc}
A & I\\
-I & B 
\end{array}
\right|
=
-\sum_{s=1}^2
\left|
\begin{array}{ccc}
A & I & {\bf 0}^T \\
-I & B & -C_s^{\dagger}\\
{\bf 0} & C_s & 0 
\end{array}
\right|
\left|
\begin{array}{ccc}
A & I & \phi \\
-I & B & {\bf 0}^T\\
-\phi^\dagger & {\bf 0} & 0
\end{array}
\right|\nonumber\\&&\hspace{-8.2cm}
+\sum_{s=1}^2
\left|
\begin{array}{ccc}
A & I & \phi\\
-I & B & {\bf 0}^T\\
{\bf 0} & C_s & 0
\end{array}
\right|
\left|
\begin{array}{ccc}
A & I & {\bf 0}^T\\
-I & B & -C_s^{\dagger}\\
-\phi^\dagger & {\bf 0} & 0
\end{array}
\right|.
\eea
\endnumparts
Thus equations (8i) and (8j) clearly show that the given Gram determinants $g^{(s)}$ and $f$ satisfy the bilinear equations (3), which completes the proof of (6) with (7).
\subsection{$(N, M, N+M)$ soliton solution}
We now point out that the $(N, M, N+M)$ soliton solution (for $N$ even) given in ref. \cite{ref13} can be obtained  as a special case of  the above $(m, m, m)$ soliton solution for the specific choice of parameters $\alpha_i^{(2)}=0, i=1, 2, \ldots, N$, and $\alpha_l^{(1)}=0, l=N+1, N+2, \ldots, m (=N+M)$ along with the parametric restrictions 
\bea
&&\alpha_i^{(1)}=\frac{\prod_{j=1}^m (k_i+k_j^*)}{\prod_{j=1,i\neq j}^m (k_j-k_i)},\quad i=1, 2, \ldots, N,\nonumber\\
&&\alpha_l^{(2)}=\frac{\prod_{j=1}^m (k_l+k_j^*)}{\prod_{j=1,l\neq j}^m (k_j-k_l)},\quad l=N+1, N+2, \ldots, m (=N+M).\nonumber
\eea
In the following sections, we will consider the explicit cases of $m=1, 2, 3$, and 4 soliton solutions and the nature of the soliton interactions therein.

\section{One-soliton (1, 1, 1) and two-soliton (2, 2, 2) solutions}
Specializing to the case of $m=1$ in equation (6) so that the Gram determinants take the form 
\bea
g^{(j)}=
\left|
\begin{array}{ccc}
A_{11} & 1 & e^{\eta_1}\\
-1 & B_{11} & 0\\
0 & -\alpha_1^{(j)} & 0
\end{array}
\right|, \quad \quad f= \left|
\begin{array}{cc}
A_{11} & 1\\
-1 & B_{11} 
\end{array}
\right|, \quad j=1,2,
\eea
\label{1-d}
\noindent where $\ds{A_{11}=\frac{e^{\eta_1+\eta_1^*}}{k_1+k_1^*}}$, and $B_{11}=\kappa_{11}=\ds{\frac{-(|\alpha_1^{(1)}|^2+|\alpha_1^{(2)}|^2)}{\omega_1+\omega_1^*}}.$
One can write down the explicit one-soliton solution as 
\numparts
\bea
&&S^{(j)}=\frac{{\alpha_1^{(j)}e^{\eta_1}}} {1+e^{{\eta_1+\eta_1^*+R}}},\quad j=1,2,\\
&&L=-2\frac{\partial^2}{\partial x^2}
\left(\log\left(1+e^{{\eta_1+\eta_1^*+R}}\right)\right),
\eea
where
\bea
\eta_1&&=k_1x-(ik_1^2+\omega_1)y+\omega_1t, \quad e^R=\frac{-\sum_{j=1}^{2}(\alpha_1^{(j)}\alpha_1^{(j)*})}{4k_{1R}\omega_{1R}},\\
k_1&&=k_{1R}+ik_{1I},\quad \omega_1=\omega_{1R}+i\omega_{1I}.
\eea
\label{genone}
\endnumparts 
Here $\alpha_1^{(1)}$, $\alpha_1^{(2)}$, $\omega_1$ and $k_1$ are all complex parameters.  In equation (10) the suffixes $R$ and $I$ denote the real and imaginary parts, respectively.  It may be noted that this bright soliton solution is nonsingular only when $k_{1R}\omega_{1R}<0$,  otherwise equation (10) becomes singular.  In this work, the main focus will be on nonsingular solutions as they are of physical importance.
The above one-soliton solution can also be rewritten as 
\numparts
\bea
&&S^{(j)}= A_j\sqrt{k_{1R}\omega_{1R}} e^{i\eta_{1I}} \mbox{sech} \left(\eta_{1R}+\frac{R}{2}\right) ,\quad j=1,2,\\
&&L=-2k_{1R}^2 \mbox{sech}^2 \left(\eta_{1R}+\frac{R}{2}\right),
\eea
where
\bea
\hspace{-2.2cm}\eta_{1R}=k_{1R}x+(2k_{1R}k_{1I}-\omega_{1R})y+\omega_{1R}t\;\; \mbox{and}\;\; A_j=\frac{\alpha_1^{(j)}}{\left(|\alpha_1^{(1)}|^2+|\alpha_1^{(2)}|^2\right)^{\frac{1}{2}}}, \quad j=1,2.\nonumber
\eea
\label{one}
\endnumparts
The complex quantities $A_j\sqrt{k_{1R}\omega_{1R}}$, $j=1,2,$ represent the amplitude of the soliton in the $S^{(j)}$ components whereas thy real quantity $2k_{1R}^2$ gives the amplitude of the soliton in the component $-L$.  Note that the complex quantities $A_1$ and $A_2$ satisfy the relation $|A_1|^2+|A_2|^2=1$, which is a reflection of the fact that the set of equation (1) is rotationally symmetric in the ($S^{(1)}, S^{(2)}$) space.  

For illustrative purpose, let us obtain the soliton solution for the special choice of parameters $\omega_1=-ik_1^2/2$.  In this case, the above soliton solution (11) becomes
\numparts
\bea
&&S^{(j)}= A_j k_{1R}\sqrt{k_{1I}} e^{i\eta_{1I}} \mbox{sech} \left(\eta_{1R}+\frac{R}{2}\right),\quad j=1,2,\label{special_s}\\
&&L=-2k_{1R}^2 \mbox{sech}^2 \left(\eta_{1R}+\frac{R}{2}\right),
\eea
where 
\bea
\eta_1=k_1x-\frac{ik_1^2}{2} (t+y), \quad e^R=\frac{\sum_{j=1}^{2}(\alpha_1^{(j)}\alpha_1^{(j)*})}{-4k_{1R}^2 k_{1I}},\quad k_1=k_{1R}+ik_{1I},
\label{suff}
\eea
\label{special}
\endnumparts
The above soliton solution is nonsingular only when $k_{1I}\leq 0$, otherwise the parameter $R$ in equation (\ref{suff}) becomes complex and the solution (12) becomes singular.  Interestingly, we observe that by just reducing the width of the soliton in the L component (which is proportional to $k_{1I}$) without affecting its amplitude, the soliton in the $S^{(1)}$ and $S^{(2)}$ components can be amplified with a proportionate pulse compression, a desirable property for a pulse in nonlinear optics.
 
\subsection{Two-soliton (2, 2, 2) solution}
To obtain the two soliton solution, we take $m=2$ in equation (7) and deduce the Gram determinant forms as
\bea
\hspace{-2.5cm}g^{(j)}=
\left|
\begin{array}{ccccc}
A_{11} & A_{12}&1&0& e^{\eta_1}\\
A_{21} & A_{22}&0&1& e^{\eta_2}\\
-1&0 & B_{11} &B_{12} & 0\\
0&-1 & B_{21} &B_{22} & 0\\
0 &0& -\alpha_1^{(j)}&-\alpha_2^{(j)} & 0
\end{array}
\right|, \quad \quad f= \left|
\begin{array}{cccc}
A_{11} &A_{12}& 1&0\\
A_{21} &A_{22}& 0&1\\
-1&0 & B_{11}&B_{12} \\
0&-1 & B_{21}&B_{22} \\
\end{array}
\right|,
\eea
\label{Det2sol}
\noindent where $A_{ij}=\ds{\frac{e^{\eta_i+\eta_j^*}}{k_i+k_j^*}}$, and $B_{ij}=\kappa_{ji}=\ds{-\frac{\left(\alpha_j^{(1)}\alpha_i^{(1)*}+\alpha_j^{(2)}\alpha_i^{(2)*}\right)}{(\omega_j+\omega_i^*)}}$, \;\;\;$i,j=1,2$.
We can then write the explicit form of the (2, 2, 2) soliton solution as
\numparts
\bea
\hspace{-2cm}S^{(j)}=\frac{1}{f}\Big(\alpha_1^{(j)} e^{\eta_1}+\alpha_2^{(j)} e^{\eta_2}+e^{\eta_1+\eta_1^{*}+\eta_2+\delta_{1j}}+e^{\eta_2+\eta_2^{*}+\eta_1+\delta_{2j}}\Big),\quad j=1,2,\\
\hspace{-2cm}L=-2\frac{\partial^2}{\partial x^2}\log(f),
\eea
where 
\bea
f=&&1+e^{\eta_1+\eta_1^{*}+R_1}+e^{\eta_1+\eta_2^{*}+\delta_0}+e^{\eta_2+\eta_1^{*}+\delta_0^{*}}+e^{\eta_2+\eta_2^{*}+R_2}\nonumber\\&&+e^{\eta_1+\eta_1^{*}+\eta_2+\eta_2^{*}+R_3}.
\eea
\label{2sol}
\endnumparts
The various quantities found in equation (14) are defined as below:
\numparts
\bea
\eta_i&&=k_ix-(ik_i^2+\omega_i)y+\omega_it,\quad i=1,2, \quad e^{R_1}=\frac{\kappa_{11}}{(k_1+k_1^*)},\\
e^{R_2}&&=\frac{\kappa_{22}}{(k_2+k_2^*)},\quad e^{\delta_0}=\frac{\kappa_{12}}{(k_1+k_2^*)}, \quad e^{\delta_0^*}=\frac{\kappa_{21}}{(k_2+k_1^*)},\\
e^{\delta_{1j}}&&=\frac{(k_1-k_2)}{(k_1+k_1^*)(k_2+k_1^*)}(\alpha_1^{(j)}\kappa_{21}-\alpha_2^{(j)}\kappa_{11}),\\
e^{\delta_{2j}}&&=\frac{(k_2-k_1)}{(k_2+k_2^*)(k_1+k_2^*)}(\alpha_2^{(j)}\kappa_{12}-\alpha_1^{(j)}\kappa_{22}),\quad j=1,2,\\
e^{R_3}&&=\frac{|k_1-k_2|^2}{(k_1+k_1^*)(k_2+k_2^*)|k_1+k_2^*|^2}(\kappa_{11}\kappa_{22}-\kappa_{12}\kappa_{21}),\\
\kappa_{il}&&=-\frac{\left(\alpha_i^{(1)}\alpha_l^{(1)*}+\alpha_i^{(2)}\alpha_l^{(2)*}\right)}{(\omega_i+\omega_l^*)}, \quad i,l=1,2.\nonumber
\eea
\label{2_asym}
The two-soliton solution is characterized by eight arbitrary complex parameters $\alpha_1^{(1)}$, $\alpha_1^{(2)}$, $\alpha_2^{(1)}$, $\alpha_2^{(2)}$, $k_1$, $k_2$, $\omega_1$ and $\omega_2$.  The above solution features both singular and nonsingular solutions.  
The nonsingular solution can be obtained by requiring the denominator $f$ in (14) to be real and nonzero.  The expression (14c) for $f$ can be rewritten as
\bea
&&\hspace{-2cm}f=2 e^{\eta_{1R}+\eta_{2R}}\left(e^{(R_1+R_2)/2} \cosh\left(\eta_{1R}-\eta_{2R}+(R_1+R_2)/2\right)+e^{\delta_{0R}} \cos\left(\eta_{1I}-\eta_{2I}+\delta_{0I}\right)\right.\nonumber\\
&&\hspace{-1cm}\left.+e^{R_3/2} \cosh\left(\eta_{1R}+\eta_{2R}+R_3/2\right)\right).
\eea
\endnumparts
To get regular solutions, $e^{R_1}$ and $e^{R_2}$ should be positive which can be obtained only for $k_{1R} \omega_{1R}<0$ and $k_{2R} \omega_{2R}<0$, respectively.  Otherwise, that is for negative values, the solution is not regular as in this case $R_1$ and $R_2$ appearing in the argument of $\cosh$ in first term become complex.  So the condition $k_{jR} \omega_{jR}<0, j=1,2$ is a necessary condition to obtain regular solution.  In a similar way, in the third term, the quantity $R_3/2$ becomes real and positive for the condition $\kappa_{11}\kappa_{22}-|\kappa_{12}|^2>0$, as may be seen from equation (15e).  Still the middle term $\cos(\eta_{1I}-\eta_{2I}+\delta_{0I})$ can lead to a singularity as it oscillates between -1 and 1.  This can be eliminated by choosing the coefficients of the remaining two terms as $e^{(R_1+R_2)/2}+e^{R_3/2}>e^{\delta_{0R}}$, in order to ensure that $f$ will not be zero at any point in space and time.  The last condition is a sufficient one.   As an illustration, the interaction of two solitons in system (1) is shown in figure \ref{inelastic}.  The parameters are chosen as $k_1=1-2i$, $k_2=1.5-1.05i$, $\omega_1=-1-i$, $\omega_2=-1.3-0.5i$, $\alpha_1^{(1)}=2$, $\alpha_2^{(1)}=\alpha_1^{(2)}=1$, $\alpha_2^{(2)}=0.01$.  One observes that the solitons in the $S^{(1)}$ and $S^{(2)}$ components undergo shape changing (energy redistribution) collisions while there is only elastic collision in the $L$ component.  More details are given in section 6 below.

\subsection{(1, 1, 2) soliton solution of Ohta \etal }
Now we show that the (1, 1, 2) soliton solution obtained by Ohta \etal \cite{ref13}  is a special case of the above two-soliton (2, 2, 2) solution (14).  Specifically, for the special choice of the parameters $\alpha_2^{(1)}=\alpha_1^{(2)}=0$, the above two-soliton solution becomes
\numparts
\bea
&&S^{(1)}=\frac{1}{f}\left(\alpha_1^{(1)} e^{\eta_1}+e^{\eta_2+\eta_2^{*}+\eta_1+\delta_{21}}\right),\\
&&S^{(2)}=\frac{1}{f}\left(\alpha_2^{(2)} e^{\eta_2}+e^{\eta_1+\eta_1^{*}+\eta_2+\delta_{12}}\right),\\
&&L=-2\frac{\partial^2}{\partial x^2}\left(\log(f)\right), 
\eea
where
\bea
f=&&1+e^{\eta_1+\eta_1^{*}+R_1}+e^{\eta_2+\eta_2^{*}+R_2}+e^{\eta_1+\eta_1^{*}+\eta_2+\eta_2^{*}+R_3}.
\eea
\label{2sol_special}
The various other parameters defined in equations (14) now take the forms
\bea
&&e^{R_1}=\frac{\kappa_{11}}{(k_1+k_1^*)}, \quad e^{R_2}=\frac{\kappa_{22}}{(k_2+k_2^*)},\quad e^{\delta_0}=e^{\delta_{11}}=e^{\delta_{22}}=0,\\
&&e^{\delta_{12}}=\frac{-\alpha_2^{(2)}\kappa_{11} (k_1-k_2)}{(k_1+k_1^*)(k_2+k_1^*)},\quad e^{\delta_{21}}=\frac{-\alpha_1^{(1)}\kappa_{22}(k_2-k_1)}{(k_2+k_2^*)(k_1+k_2^*)},\\
&&e^{R_3}=\frac{|k_1-k_2|^2 \kappa_{11}\kappa_{22}}{(k_1+k_1^*)(k_2+k_2^*)|k_1+k_2^*|^2},\\
&&\kappa_{11}=-\frac{|\alpha_1^{(1)}|^2}{(\omega_1+\omega_1^*)}, \quad \kappa_{22}=-\frac{|\alpha_2^{(2)}|^2}{(\omega_2+\omega_2^*)}.
\eea
\label{special_2}
Solution (16a-h) is nothing but the (1, 1, 2) soliton solution obtained by Ohta \etal in ref. \cite{ref13} when the parameters in (16a-h) are further restricted to the special choice
\bea
\alpha_1^{(1)}=\frac{(k_1+k_1^*)(k_1+k_2^*)}{(k_2-k_1)}\;\;\;\mbox{and}\;\;\; \alpha_2^{(2)}=\frac{(k_2+k_2^*)(k_2+k_1^*)}{(k_2-k_1)}.
\eea
\endnumparts
\section{Asymptotic analysis of the two soliton solution (14) of the three wave system}
We now consider the collision properties associated with the general two-soliton solution (14) of the three wave system.  For this purpose we carry out the analysis, for $k_{jR}>0$, $\omega_{jR}<0, j=1,2$.   Also we choose $\frac{k_{2R}}{k_{1R}}>\left|\frac{\omega_{2R}}{\omega_{1R}}\right|$ and $\frac{k_{2R}k_{2I}}{k_{1R}k_{1I}}>\left|\frac{\omega_{2R}}{\omega_{1R}}\right|$ for convenience.  Similar analysis can be performed for other choices of $k_{jR}$'s and $\omega_{jR}$'s also by keeping $k_{jR}>0$, $\omega_{jR}<0$, which is the necessary condition for nonsingular solutions.   We now define the soliton wave variables as $\eta_{1R}=k_{1R}x+(2k_{1R}k_{1I}-\omega_{1R})y+\omega_{1R}t$ and $\eta_{2R}=k_{2R}x+(2k_{2R}k_{2I}-\omega_{2R})y+\omega_{2R}t$.  In the limit $x, y \rightarrow \pm \infty$ and a fixed $t$ the two-soliton solution (14) takes the following asymptotic forms.

\noindent\underline {{\bf {a)}} Before collision (limit $x, y\rightarrow-\infty$):}\\ 
(i)  Soliton 1 ($\eta_{1R}\simeq0, \eta_{2R}\rightarrow -\infty$):
\numparts
\bea
\left(
\begin{array}{c}
S^{(1)} \\\\
 S^{(2)}
\end{array}
\right)\simeq \left(
\begin{array}{c}
 A_{1}^{1-} \\\\
 A_{2}^{1-}  \\ 
\end{array}
\right) \sqrt{k_{1R}\omega_{1R}}\;\;\mbox{\mbox{sech}}\left(\eta_{1R}+\frac{R_1}{2}
\right)e^{i\eta_{1I}},\\
L\simeq -2 k_{1R}^2\;\; \mbox{sech}^2 \left(\eta_{1R}+\frac{R_1}{2}
\right),
\eea
where
\bea
\left(
\begin{array}{c}
A_{1}^{1-} \\\\
A_{2}^{1-}
\end{array}
\right)\simeq \left(
\begin{array}{c}
\alpha_1^{(1)}\\\\
\alpha_1^{(2)}
\end{array}
\right)  \frac{e^{-R_1/2}}{((k_1+k_1^*)(\omega_1+\omega_1^*))^{1/2}}.
\label{A1b}
\eea
\label{bs1}
\endnumparts
(ii)  Soliton 2 ($\eta_{2R}\simeq0, \eta_{1R}\rightarrow \infty$):
\numparts
\bea
\left(
\begin{array}{c}
 S^{(1)} \\\\
 S^{(2)}
\end{array}
\right)\simeq \left(
\begin{array}{c}
 A_{1}^{2-} \\\\
 A_{2}^{2-} 
\end{array}
\right) \sqrt{k_{2R}\omega_{2R}}\;\; \mbox{\mbox{sech}}\left(\eta_{2R}+\frac{(R_3-R_1)}{2}
\right)e^{i\eta_{2I}},\\
L\simeq  -2 k_{2R}^2\;\; \mbox{sech}^2 \left(\eta_{2R}+\frac{(R_3-R_1)}{2}
\right),
\eea
where
\bea
\left(
\begin{array}{c}
A_{1}^{2-} \\\\
A_{2}^{2-}
\end{array}
\right)\simeq \left(
\begin{array}{c}
e^{\delta_{11}}\\\\
e^{\delta_{12}}
\end{array}
\right)\frac{e^{-(R_1+R_3)/2}}{((k_2+k_2^*)(\omega_2+\omega_2^*))^{1/2}}.
\label{A2b}
\eea
\label{bs2}
\endnumparts
\begin{flushleft} The various quantities in the above expressions are defined in equation (15).\end{flushleft}

\noindent\underline {{\bf {b)}} After collision (limit $x, y\rightarrow\infty$):}\\ 
(i)  Soliton 1 ($\eta_{1R}\simeq0, \eta_{2R}\rightarrow \infty$):
\numparts
\bea
\left(
\begin{array}{c}
S^{(1)} \\\\
S^{(2)} 
\end{array}
\right)\simeq \left(
\begin{array}{c}
 A_{1}^{1+} \\\\
 A_{2}^{1+} 
\end{array}
\right)\sqrt{k_{1R}\omega_{1R}}\;\;\mbox{\mbox{sech}}\left(\eta_{1R}+\frac{(R_3-R_2)}{2}
\right)e^{i\eta_{1I}},\\
L\simeq -2 k_{1R}^2\;\;\mbox{sech}^2\left(\eta_{1R}+\frac{(R_3-R_2)}{2}
\right),
\label{2aq3-1}
\eea
where
\bea
\left(
\begin{array}{c}
A_{1}^{1+} \\\\
A_{2}^{1+}
\end{array}
\right)\simeq \left(
\begin{array}{c}
e^{\delta_{21}}\\\\
e^{\delta_{22}}
\end{array}
\right)\frac{e^{-(R_2+R_3)/2}}{((k_1+k_1^*)(\omega_1+\omega_1^*))^{1/2}}.
\label{A1a}
\eea
\label{as1}
\endnumparts
(ii)  Soliton 2 ($\eta_{2R}\simeq0, \eta_{1R}\rightarrow -\infty$):
\numparts
\bea
\left(
\begin{array}{c}
S^{(1)} \\\\
S^{(2)}
\end{array}
\right)\simeq \left(
\begin{array}{c}
 A_{1}^{2+} \\\\
 A_{2}^{2+} 
\end{array}
\right)\sqrt{k_{2R}\omega_{2R}}\;\;\mbox{\mbox{sech}}\left(\eta_{2R}+\frac{R_2}{2}
\right)e^{i\eta_{2I}},\\
L\simeq -2 k_{2R}^2\;\;\mbox{sech}^2\left(\eta_{2R}+\frac{R_2}{2}
\right),
\label{2aq3-2}
\eea
where
\bea
\left(
\begin{array}{c}
A_{1}^{2+} \\\\
A_{2}^{2+}
\end{array}
\right)\simeq \left(
\begin{array}{c}
\alpha_2^{(1)} \\\\
\alpha_2^{(2)}
\end{array}
\right) \frac{e^{-R_2/2}}{((k_2+k_2^*)(\omega_2+\omega_2^*))^{1/2}}.
\label{A2a}
\eea
\label{as2}
\endnumparts
Note that in all the above expressions $\left|A_1^{j\pm}\right|^2+\left|A_2^{j\pm}\right|^2=1$, $j=1,2$.  Our above analysis reveals the fact that due to collision the amplitude of the colliding solitons, say $s_1$ and $s_2$ in the $S^{(1)}$ and $S^{(2)}$ components, change from $\left(A_1^{1-}, A_2^{1-}\right)\sqrt{k_{1R}\omega_{1R}}$\; and $\left(A_1^{2-}, A_2^{2-}\right) \sqrt{k_{2R}\omega_{2R}}$ to $\left(A_1^{1+}, A_2^{1+}\right) \sqrt{k_{1R}\omega_{1R}}$\; and $\left(A_1^{2+}, A_2^{2+}\right) \sqrt{k_{2R}\omega_{2R}}$, respectively.  Here the superscripts in $A_i^{j\pm}$'s with $i, j=1,2$ denote the solitons $s_1$ and $s_2$, while the subscripts represent the components $S^{(1)}$ and $S^{(2)}$ and the $``\pm"$ signs stand for $``x, y\rightarrow\pm\infty"$.  In addition to this change in the amplitudes, the solitons also undergo amplitude-dependent phase shifts due to the collision and they can be determined straightforwardly from the above asymptotic expressions.  From equations ((17) and (19)) and equations ((18) and (20)), one can easily check that the phase shift suffered by the soliton $s_1$ (say $\Phi_1)=-$ Phase shift of soliton $s_2$ (say $-\Phi_2\equiv \Phi_1$) = $\Phi$ and is given by
\bea
\Phi=\frac{(R_3-R_1-R_2)}{2},
\label{p}
\eea
where $R_1$, $R_2$ and $R_3$ are as defined in equation (15) and depends on the amplitudes.

\section{Soliton Interaction}
\begin{figure}
\centering
\includegraphics[width=0.6\linewidth]{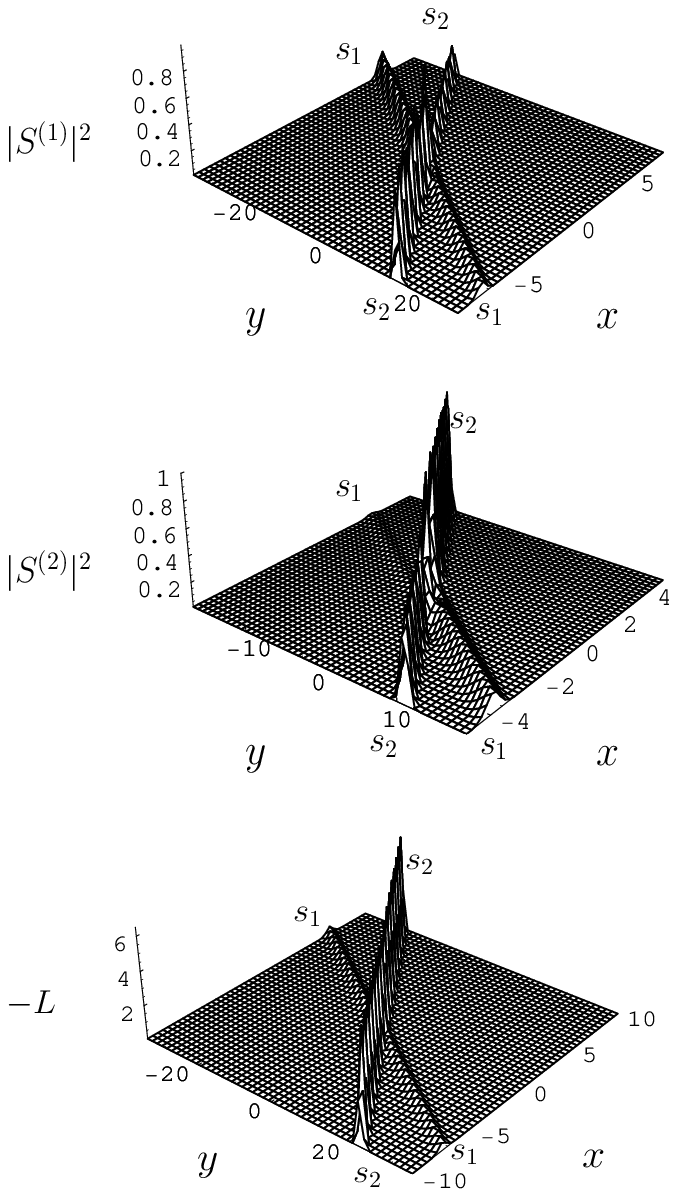}
\caption{Shape changing collision of solitons in the three wave system with the parametric choice $\frac{\alpha_1^{(1)}}{\alpha_2^{(1)}}\neq\frac{\alpha_1^{(2)}}{\alpha_2^{(2)}}$.  Here the two soliton solution (14) is plotted for a fixed value of $t$ and the associated parameters are given in the text (below equation (15)).  Note that intensity redistribution occurs only in the $S^{(1)}$ and $S^{(2)}$ components, while elastic collision only occurs in the $L$ component.}
\label{inelastic}
\end{figure}

Now it is of further interest to analyze the interaction properties of the solitons depicted in figure \ref{inelastic} for the specific set of values of the parameters given in section 4.1.  Figure \ref{inelastic}  shows typical spatial collision of two solitons for $t=-4$ corresponding to the exact expression (14).  The interesting collision scenario  depicted in figure \ref{inelastic} clearly indicates that there is a redistribution of intensity among the two $S^{(j)}$ components resulting in an enhancement (suppression) of intensity of solitons $s_2$ ($s_1$) in the $S^{(1)}$ component and a suppression (enhancement) of soliton $s_2$ ($s_1$) in the $S^{(2)}$ component.  The solitons also undergo amplitude-dependent phase shifts along with this energy redistribution.  However, the solitons appearing in the long wave component ($L$) exhibit the standard elastic collision as shown in the third figure of figure \ref{inelastic} though the phase shift here is also amplitude-dependent.  Interestingly, if the parameters are so chosen such that the condition $\frac{\alpha_1^{(1)}}{\alpha_2^{(1)}}=\frac{\alpha_1^{(2)}}{\alpha_2^{(2)}}$ is satisfied, there occurs only elastic collision in all the three components $S^{(j)}$ and $L$.  The underlying collision dynamics can be well understood by using the asymptotic analysis of the two-soliton solution (14) discussed in section 5 and is further described below. 
\subsection{Collision behaviour of solitons in the short wave components}
The asymptotic analysis presented in the previous section also results in the following expressions relating the intensities of solitons $s_1$ and $s_2$ in the $S^{(1)}$ and $S^{(2)}$ components before and after interaction,
\numparts
\bea
|A_i^{j+}|^2=|T_j^{i}|^2 |A_i^{j-}|^2, \quad i,j=1,2,
\eea
where the superscripts $j\pm$ represent the solitons designated as $s_1$ and $s_2$ at $``x, y\rightarrow\pm\infty"$.  The expression for the transition intensities for the solitons in the short wave components can be written down using the results in equations (\ref{A1b}), (\ref{A2b}), (\ref{A1a}) and (\ref{A2a}) as    
\bea
|T_j^{1}|^2&=&\frac{|1-\lambda_2(\alpha_2^{(j)}/\alpha_1^{(j)})|^2}{|1-\lambda_1\lambda_2|},\\
|T_j^{2}|^2&=&\frac{|1-\lambda_1\lambda_2|}{|1-\lambda_1(\alpha_1^{(j)}/\alpha_2^{(j)})|^2}, \quad j=1,2,\\
\lambda_1&=&\frac{\kappa_{21}}{\kappa_{11}}, \quad \lambda_2=\frac{\kappa_{12}}{\kappa_{22}}.
\eea
\label{trans}
\endnumparts
In general $\left|T_j^i\right|\neq 1$ and so an intensity (energy) redistribution of the solitons in the $S^{(1)}$ and $S^{(2)}$ components occurs as shown in figure \ref{inelastic}.  One can notice that the standard elastic collision takes place for the specific parametric choice $\frac{\alpha_1^{(1)}}{\alpha_2^{(1)}}=\frac{\alpha_1^{(2)}}{\alpha_2^{(2)}}$, as $|T_j^i|^2=1$ and hence $|A_i^{j-}|^2=|A_i^{j+}|^2$, $i,j=1,2,$ for this choice.  However the two colliding solitons $s_1$ and $s_2$ suffer amplitude-dependent phase shifts $\Phi_1$ and $\Phi_2$, respectively, as given in equation (\ref{p}).
\subsection{Collision scenario in the long wave component}
 In the $L$ component, there occurs only elastic collision for any parametric choice. This is evident from the asymptotic analysis, vide equations (17b), (18b), (19b) and (20b).  One finds that the amplitudes of the solitons $s_1$ and $s_2$ before and after interaction are the same which are $-2k_{1R}^2$ and $-2k_{2R}^2$, respectively, while there occurs an amplitude-dependent phase shift as given by equation (\ref{p}).
\subsection{Shape changing collisions and Linear fractional transformations}
It is instructive to notice that the intensity redistribution in the short wave components  characterized by the transition matrices (equation (22)) can also be viewed as a linear fractional transformation (LFT).  To realize this, we re-express the amplitude changes in the short wave components of soliton $s_1$ after interaction as
\numparts
\bea
A_1^{1+}=\Gamma C_{11} A_1^{1-}+\Gamma C_{12} A_2^{1-},\label{lft1}\\
A_2^{1+}=\Gamma C_{21} A_1^{1-}+\Gamma C_{22} A_2^{1-}.
\label{lft2}
\eea
Here
\bea
&&\Gamma=\left(\frac{a}{a^*}\right)c\left[(\alpha_1^{(1)}\alpha_2^{(1)*}+\alpha_1^{(2)}\alpha_2^{(2)*})(\alpha_2^{(1)}\alpha_2^{(1)*}+\alpha_2^{(2)}\alpha_2^{(2)*})\right]^{-1},\\
&&C_{11}=-\left[(\alpha_2^{(1)}\alpha_2^{(1)*})(\omega_1-\omega_2)+(\alpha_2^{(2)}\alpha_2^{(2)*})(\omega_1+\omega_2^*)\right],\\
&&C_{12}=(\alpha_2^{(1)}\alpha_2^{(2)*})(\omega_2+\omega_2^*),\\
&&C_{21}=(\alpha_2^{(2)}\alpha_2^{(1)*})(\omega_2+\omega_2^*),\\
&&C_{22}=-\left[(\alpha_2^{(1)}\alpha_2^{(1)*})(\omega_1+\omega_2^*)+(\alpha_2^{(2)}\alpha_2^{(2)*})(\omega_1-\omega_2)\right],
\eea
where
\bea
&&c=\left(\frac{1}{|\kappa_{12}|^2}-\frac{1}{\kappa_{11}\kappa_{22}}\right)^{-1/2},\\
&&a=\left[-(k_1-k_2)(k_2+k_1^*)(\omega_2+\omega_1^*)(\alpha_1^{(1)}\alpha_2^{(1)*}+\alpha_1^{(2)}\alpha_2^{(2)*})\right]^{1/2}.
\eea
Note that the coefficients $C_{ij}$'s, $i,j=1,2,$ are independent of $\alpha_1^{(j)}$'s and so of $A_1^{1-}$ and $A_2^{1-}$, that is the $\alpha$ parameters of soliton $s_1$.  From equations (\ref{lft1}) and (\ref{lft2}),
\bea
\rho_{1,2}^{1+}=\frac{A_1^{1+}}{A_2^{1+}}=\frac{C_{11}\rho_{1,2}^{1-}+C_{12}}{C_{21}\rho_{1,2}^{1-}+C_{22}},
\eea
\endnumparts
where $\rho_{1,2}^{1-}=\frac{A_1^{1-}}{A_2^{1-}}$, in which the superscripts represent the underlying soliton and the subscripts represent the corresponding short wave components.  Thus the state of $s_1$ before and after interaction is characterized by the complex quantities $\rho_{1,2}^{1-}$ and $\rho_{1,2}^{1+}$, respectively.  The direct consequence of the above LFT representation is the identification of a binary logic using soliton collisions as in the case of CNLS equations \cite{ref5,ref11,ref12} and hence the LFT can be profitably used to construct logic gates associated with the binary logic.  A similar analysis can be made for the soliton $s_2$ also. 
\section{ Three-soliton (3, 3, 3) solution}
\begin{figure}
\centering
\includegraphics[width=0.6\linewidth]{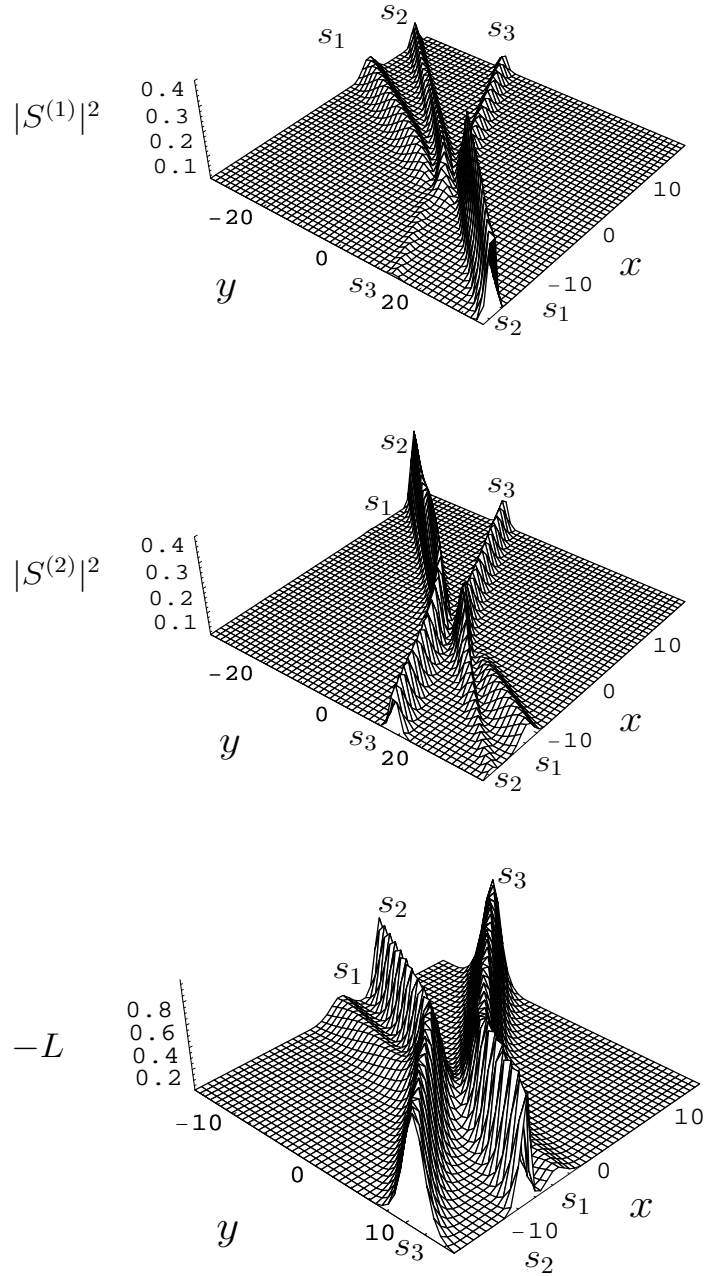}
\caption{Shape changing collision of the three solitons of the three wave system, equation (24).  The chosen soliton parameters are given in the text.  Again note that intensity redistribution occurs only in the $S^{(1)}$ and $S^{(2)}$ components, while elastic collision only occurs in the $L$ component.}
\label{three}
\end{figure}

From the general form (7), and restricting $m=3$, one can write down the explicit three-soliton (3, 3, 3) solution as 
\numparts
\bea
S^{(j)}&=& \frac{\alpha_1^{(j)}e^{\eta_1}+\alpha_2^{(j)}e^{\eta_2}+\alpha_3^{(j)}
e^{\eta_3}
+e^{\eta_1+\eta_1^*+\eta_2+\delta_{1j}}+e^{\eta_1+\eta_1^*+\eta_3+\delta_{2j}}
+e^{\eta_2+\eta_2^*+\eta_1+\delta_{3j}}}
{f}\nonumber\\
&&+\frac{e^{\eta_2+\eta_2^*+\eta_3+\delta_{4j}}
+e^{\eta_3+\eta_3^*+\eta_1+\delta_{5j}}+e^{\eta_3+\eta_3^*+\eta_2+\delta_{6j}}
+e^{\eta_1^*+\eta_2+\eta_3+\delta_{7j}}
+e^{\eta_1+\eta_2^*+\eta_3+\delta_{8j}}
}{f} \nonumber\\
 &&+\frac{e^{\eta_1+\eta_2+\eta_3^*+\delta_{9j}}
+e^{\eta_1+\eta_1^*+\eta_2+\eta_2^*+\eta_3+\tau_{1j}}
+e^{\eta_1+\eta_1^*+\eta_3+\eta_3^*+\eta_2+\tau_{2j}}}{f}\nonumber\\
&&+\frac{
e^{\eta_2+\eta_2^*+\eta_3+\eta_3^*+\eta_1+\tau_{3j}}}{f},\;\;j=1,2,
\eea
where
\bea
f &=&1+e^{\eta_1+\eta_1^*+R_1}+e^{\eta_2+\eta_2^*+R_2}+e^{\eta_3+\eta_3^*+R_3}
+e^{\eta_1+\eta_2^*+\del_{10}}+e^{\eta_1^*+\eta_2+\del_{10}^*}\nonumber\\
&&+e^{\eta_1+\eta_3^*+\del_{20}}
+e^{\eta_1^*+\eta_3+\del_{20}^*}
+e^{\eta_2+\eta_3^*+\del_{30}}
+e^{\eta_2^*+\eta_3+\del_{30}^*}
+e^{\eta_1+\eta_1^*+\eta_2+\eta_2^*+R_4}\nonumber\\
&&+e^{\eta_1+\eta_1^*+\eta_3+\eta_3^*+R_5}
+e^{\eta_2+\eta_2^*+\eta_3+\eta_3^*+R_6}
+e^{\eta_1+\eta_1^*+\eta_2+\eta_3^*+\tau_{10}}
+e^{\eta_1+\eta_1^*+\eta_3+\eta_2^*+\tau_{10}^*}\nonumber\\
&&+e^{\eta_2+\eta_2^*+\eta_1+\eta_3^*+\tau_{20}}
+e^{\eta_2+\eta_2^*+\eta_1^*+\eta_3+\tau_{20}^*}
+e^{\eta_3+\eta_3^*+\eta_1+\eta_2^*+\tau_{30}}
+e^{\eta_3+\eta_3^*+\eta_1^*+\eta_2+\tau_{30}^*}\nonumber\\
&&+e^{\eta_1+\eta_1^*+\eta_2+\eta_2^*+\eta_3+\eta_3^*+R_7}.
\label{3sold}
\eea
Here
\bea
\eta_i&=&k_ix-(ik_i^2+\omega_i)y+\omega_it, i=1,2,3,\\
e^{\delta_{1j}}&=&\frac{(k_1-k_2)(\al_1^{(j)}\kappa_{21}-\al_2^{(j)}\kappa_{11}
)}{(k_1+k_1^*)(k_1^*+k_2)},\;\;
e^{\delta_{2j}}=\frac{(k_1-k_3)(\al_1^{(j)}\kappa_{31}-\al_3^{(j)}\kappa_{11}
)}{(k_1+k_1^*)(k_1^*+k_3)},\nonumber\\
e^{\delta_{3j}}&=&\frac{(k_1-k_2)(\al_1^{(j)}\kappa_{22}-\al_2^{(j)}\kappa_{12}
)}{(k_1+k_2^*)(k_2+k_2^*)},\;\;
e^{\delta_{4j}}=\frac{(k_2-k_3)(\al_2^{(j)}\kappa_{32}-\al_3^{(j)}\kappa_{22}
)}{(k_2+k_2^*)(k_2^*+k_3)},\nonumber\\
e^{\delta_{5j}}&=&\frac{(k_1-k_3)(\al_1^{(j)}\kappa_{33}-\al_3^{(j)}\kappa_{13}
)}{(k_3+k_3^*)(k_3^*+k_1)},\;\;
e^{\delta_{6j}}=\frac{(k_2-k_3)(\al_2^{(j)}\kappa_{33}-\al_3^{(j)}\kappa_{23}
)}{(k_3^*+k_2)(k_3^*+k_3)},\nonumber\\
e^{\delta_{7j}}&=&\frac{(k_2-k_3)(\al_2^{(j)}\kappa_{31}-\al_3^{(j)}\kappa_{21}
)}{(k_1^*+k_2)(k_1^*+k_3)},\;\;
e^{\delta_{8j}}=\frac{(k_1-k_3)(\al_1^{(j)}\kappa_{32}-\al_3^{(j)}\kappa_{12}
)}{(k_1+k_2^*)(k_2^*+k_3)},\nonumber\\
e^{\delta_{9j}}&=&\frac{(k_1-k_2)(\al_1^{(j)}\kappa_{23}-\al_2^{(j)}\kappa_{13}
)}{(k_1+k_3^*)(k_2+k_3^*)},\nonumber
\eea
\bea
e^{\tau_{1j}}&=&\frac{(k_2-k_1)(k_3-k_1)(k_3-k_2)(k_2^*-k_1^*)}
{(k_1^*+k_1)(k_1^*+k_2)(k_1^*+k_3)(k_2^*+k_1)(k_2^*+k_2)(k_2^*+k_3)}\nonumber\\
&&\times
\left[\al_1^{(j)}(\kappa_{21}\kappa_{32}-\kappa_{22}\kappa_{31})
+\al_2^{(j)}(\kappa_{12}\kappa_{31}-\kappa_{32}\kappa_{11})
+\al_3^{(j)}(\kappa_{11}\kappa_{22}-\kappa_{12}\kappa_{21})
\right],\nonumber\\
e^{\tau_{2j}}&=&\frac{(k_2-k_1)(k_3-k_1)(k_3-k_2)(k_3^*-k_1^*)}
{(k_1^*+k_1)(k_1^*+k_2)(k_1^*+k_3)(k_3^*+k_1)(k_3^*+k_2)(k_3^*+k_3)}\nonumber\\
&&\times
\left[\al_1^{(j)}(\kappa_{33}\kappa_{21}-\kappa_{31}\kappa_{23})
+\al_2^{(j)}(\kappa_{31}\kappa_{13}-\kappa_{11}\kappa_{33})
+\al_3^{(j)}(\kappa_{23}\kappa_{11}-\kappa_{13}\kappa_{21})
\right],\nonumber\\
e^{\tau_{3j}}&=&\frac{(k_2-k_1)(k_3-k_1)(k_3-k_2)(k_3^*-k_2^*)}
{(k_2^*+k_1)(k_2^*+k_2)(k_2^*+k_3)(k_3^*+k_1)(k_3^*+k_2)(k_3^*+k_3)}\nonumber\\
&&\times
\left[\al_1^{(j)}(\kappa_{22}\kappa_{33}-\kappa_{23}\kappa_{32})
+\al_2^{(j)}(\kappa_{13}\kappa_{32}-\kappa_{33}\kappa_{12})
+\al_3^{(j)}(\kappa_{12}\kappa_{23}-\kappa_{22}\kappa_{13})
\right],\nonumber\\
\eea
\bea
\hspace{-2.4cm}e^{R_m}=\frac{\kappa_{mm}}{k_m+k_m^*}, \;\;m=1,2,3,\;\;
e^{\del_{10}}=\frac{\kappa_{12}}{k_1+k_2^*},\;\;
e^{\del_{20}}=\frac{\kappa_{13}}{k_1+k_3^*},\;\;
e^{\del_{30}}=\frac{\kappa_{23}}{k_2+k_3^*},\nonumber
\eea
\bea
e^{R_4}&=&\frac{(k_2-k_1)(k_2^*-k_1^*)}
{(k_1^*+k_1)(k_1^*+k_2)(k_1+k_2^*)(k_2^*+k_2)}
\left[\kappa_{11}\kappa_{22}-\kappa_{12}\kappa_{21}\right],\nonumber\\
e^{R_5}&=&\frac{(k_3-k_1)(k_3^*-k_1^*)}
{(k_1^*+k_1)(k_1^*+k_3)(k_3^*+k_1)(k_3^*+k_3)}
\left[\kappa_{33}\kappa_{11}-\kappa_{13}\kappa_{31}\right],\nonumber\\
e^{R_6}&=&\frac{(k_3-k_2)(k_3^*-k_2^*)}
{(k_2^*+k_2)(k_2^*+k_3)(k_3^*+k_2)(k_3+k_3^*)}
\left[\kappa_{22}\kappa_{33}-\kappa_{23}\kappa_{32}\right],\nonumber\\
e^{\tau_{10}}&=&\frac{(k_2-k_1)(k_3^*-k_1^*)}
{(k_1^*+k_1)(k_1^*+k_2)(k_3^*+k_1)(k_3^*+k_2)}
\left[\kappa_{11}\kappa_{23}-\kappa_{21}\kappa_{13}\right],\nonumber\\
e^{\tau_{20}}&=&\frac{(k_1-k_2)(k_3^*-k_2^*)}
{(k_2^*+k_1)(k_2^*+k_2)(k_3^*+k_1)(k_3^*+k_2)}
\left[\kappa_{22}\kappa_{13}-\kappa_{12}\kappa_{23}\right],\nonumber\\
e^{\tau_{30}}&=&\frac{(k_3-k_1)(k_3^*-k_2^*)}
{(k_2^*+k_1)(k_2^*+k_3)(k_3^*+k_1)(k_3^*+k_3)}
\left[\kappa_{33}\kappa_{12}-\kappa_{13}\kappa_{32}\right],\nonumber\\
e^{R_7}&=& \frac{|k_1-k_2|^2|k_2-k_3|^2|k_3-k_1|^2}
{(k_1+k_1^*)(k_2+k_2^*)(k_3+k_3^*)|k_1+k_2^*|^2|k_2+k_3^*|^2|k_3+k_1^*|^2}
\nonumber\\
&&\times\left[(\kappa_{11}\kappa_{22}\kappa_{33}-
\kappa_{11}\kappa_{23}\kappa_{32})
+(\kappa_{12}\kappa_{23}\kappa_{31}-
\kappa_{12}\kappa_{21}\kappa_{33})\right .\nonumber\\
&&\left.+(\kappa_{21}\kappa_{13}\kappa_{32}-
\kappa_{22}\kappa_{13}\kappa_{31})\right],
\eea
and
\bea
\kappa_{il}= -\frac{\sum_{n=1}^2\alpha_i^{(n)}\alpha_l^{(n)*}}
{\left(\omega_i+\omega_l^*\right)},\;i,l=1,2,3.
\end{eqnarray} 
\label{3sol}
The explicit form of $L$ can be obtained by substituting the expression for $f$ in $L=-2 \ds{\frac{\partial}{\partial x^2}(\log f)}$.  Here $\alpha_1^{(1)}$, $\alpha_2^{(1)}$, $\alpha_3^{(1)}$, $\alpha_1^{(2)}$, $\alpha_2^{(2)}$, $\alpha_3^{(2)}$, $k_1, k_2, k_3$, $\omega_1$, $\omega_2$, and $\omega_3$ are the twelve complex parameters which characterize the above three-soliton solution.  Following the arguments of ref. \cite{ref16} (see equations (28) and (29) there) and the discussion in section 4 one can show that the necessary conditions for nonsingular solution are 
\bea
\hspace{-1cm}e^{R_i}>0,\quad i=1,2, \ldots 7,
\eea
which is automatically taken care by the choice $k_{jR}\omega_{jR}<0, j=1, 2, 3$.  One can also easily show that the inequality
\bea
\hspace{-2.6cm}e^{(R_1+R_6)/2}, e^{(R_2+R_5)/2}, e^{(R_3+R_4)/2}, e^{(R_7)/2}> 4 \mbox{max} \left(e^{(\delta_{10R}+\tau_{30R})}, e^{(\delta_{20R}+\tau_{20R})}, e^{(\delta_{30R}+\tau_{10R})}\right)
\eea
\endnumparts
is the sufficient condition in order to ensure that the solution is regular. Typical shape changing collision of the three-soliton solution is shown in figure \ref{three} for the parametric choices $k_1=0.2+0.3i$, $k_2=0.6+0.4i$, $k_3=0.7+0.2i$, $\omega_1=-0.5+0.4i$, $\omega_2=-0.7+0.1i$, $\omega_3=-0.3+0.3i$, $\alpha_1^{(1)}=0.5-i$, $\alpha_2^{(1)}=0.5+i$, $\alpha_3^{(1)}=0.3+0.2i$, $\alpha_1^{(2)}=0.39+0.2i$, and $\alpha_2^{(2)}=\alpha_3^{(2)}=1$.  The above three soliton solution (24) represents the interaction of three solitons and their collision scenario can be well understood by making an asymptotic analysis following the procedure given in section 5 for the two soliton solution.  

We have identified from the asymptotic analysis that for the three interacting solitons (say, $s_1, s_2$ and $s_3$), as in the case of CNLS equations \cite{ref5,ref16}, the total transition amplitude of a particular soliton (say $s_1$) can be expressed as the product of two transition amplitudes which result respectively during the first collision of $s_1$ with $s_2$ and during the collision of the outcoming  soliton (say $s_1'$) with soliton $s_3$.  In a similar manner the net phase shift acquired by a particular soliton (say $s_1$) during the complete collision process is equal to the addition of phase shifts experienced by that soliton during its cascaded collisions with $s_2$ and $s_3$, respectively.  Thus the analysis clearly shows that the multi-soliton collision process indeed occurs in a pair-wise manner in the multicomponent (2+1)D LSRI system and there exists no multiparticle effects.  The details are similar to the CNLS system \cite{ref5,ref16} and so we do not present them here.  

\section{Four-soliton (4, 4, 4) solution } 
Again to obtain the explicit four-soliton solution, we substitute $m=4$ in the Gram determinant form (7) and obtain an expression involving exponentials.  Since it is too lengthy, we do not present the explicit form here.  However we note that the four-soliton solution is characterized by sixteen complex parameters, $k_j$, $\omega_j$, $\alpha_j^{(1)}$, $\alpha_j^{(2)}$, $j=1, 2, 3, 4$.  The nonsingular solution results for the choice $k_{jR}\omega_{jR}<0,\;\;j=1,2,3,4$.  One can check that $k_{jR}\omega_{jR}<0, \;\;j=1,2,3,4,$ are the necessary conditions for the existence of nonsingular solution and the sufficient condition can be obtained following the procedure mentioned in sections 4 and 7.
\endnumparts
\subsection{(2, 2, 4) soliton solution of Ohta \etal}
\begin{figure}
\centering
\includegraphics[width=0.6\linewidth]{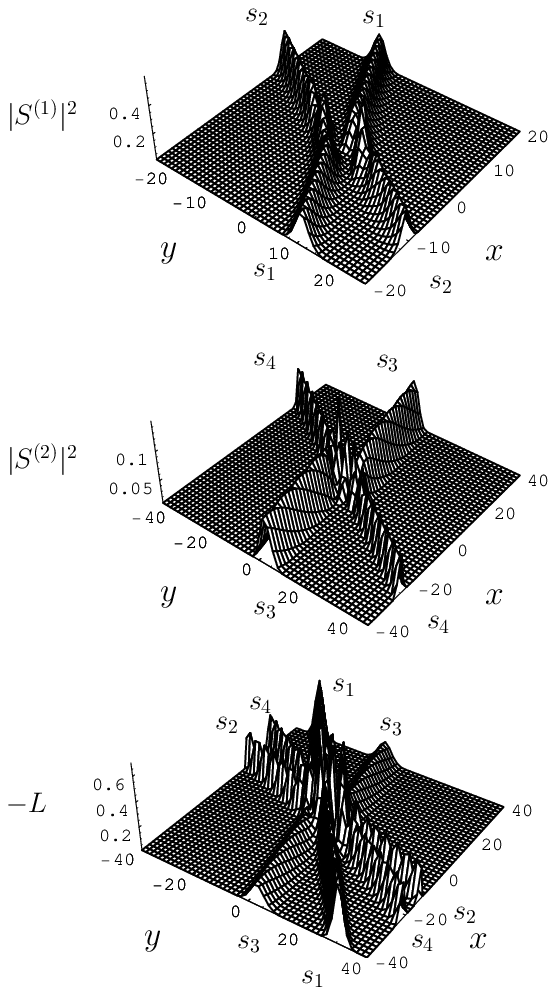}
\caption{A special case of the four soliton collision with two solitons in $S^{(1)}$ and $S^{(2)}$ components and four solitons in $L$ component, equation (25).  Note that this collision scenario is similar to the (2, 2, 4) soliton interaction depicted in figure 2 of ref. \cite{ref13}. }
\label{ohta_4}
\end{figure}

In the above discussed four-soliton solution, we make the choice $\alpha_3^{(1)}=\alpha_4^{(1)}=\alpha_1^{(2)}=\alpha_2^{(2)}=0,$ and also introduce the parametric restrictions
\numparts
\bea
&&\alpha_1^{(1)}=\frac{(k_1+k_1^*)(k_1+k_2^*)(k_1+k_3^*)(k_1+k_4^*)}{(k_2-k_1)(k_3-k_1)(k_4-k_1)},\\
&&\alpha_2^{(1)}=\frac{(k_2+k_2^*)(k_2+k_1^*)(k_2+k_3^*)(k_2+k_4^*)}{(k_1-k_2)(k_3-k_2)(k_4-k_2)},\\
&&\alpha_3^{(2)}=\frac{(k_3+k_3^*)(k_3+k_4^*)(k_3+k_2^*)(k_3+k_1^*)}{(k_4-k_3)(k_2-k_3)(k_1-k_3)},\\
&&\alpha_4^{(2)}=\frac{(k_4+k_4^*)(k_4+k_1^*)(k_4+k_2^*)(k_4+k_3^*)}{(k_3-k_4)(k_2-k_4)(k_1-k_4)},
\eea
\label{4sol_special}
\endnumparts
\hspace{-0.15cm}then one can show that it is exactly equivalent to the (2, 2, 4) soliton expression given by Ohta \etal \cite{ref13}.  This can be verified by expanding the determinant form with the above parametric restrictions and comparing it with the expanded version of the $(2, 2, 4)$ solution of ref. \cite{ref13}.  The interaction of solitons for the above special case of the four-soliton solution is shown in figure \ref{ohta_4} for the choice of parameters $k_1=0.5-0.2i$, $k_2=0.4+ 0.1i$, $k_3 = 0.3-0.4i$, $k_4=0.4+0.6i$, $\omega_1=-0.5+0.4i$, $\omega_2=-0.7 +0.1i$, $\omega_3=-0.3+0.3i$, $\omega_4=-0.2+0.2i$.  This is similar to the interaction shown in ref. \cite{ref13}.
\section{Soliton solutions of $(n+1)$-wave system}
We now extend our study to obtain multisoliton solutions of the multicomponent system with arbitrary $(n+1)$ waves, in which we consider $n$ short wave components and single long wave component.  The $(n+1)$-wave system in this case is given by
\numparts  
\bea
&&i(S_t^{(j)}+S_y^{(j)})-S_{xx}^{(j)}+L S^{(j)}=0, \quad j=1,2,\ldots, n,\\
&&L_t=2\sum_{j=1}^{n} |S^{(j)}|^2_x.
\eea
\label{n-1}
\endnumparts
\underline{(i) One soliton solution:}\\
Following the procedure discussed in section 4, we can obtain the one soliton solution as
\numparts
\bea
S^{(j)}=&&\frac{{\alpha_1^{(j)}e^{\eta_1}}} {1+e^{{\eta_1+\eta_1^*+R}}},\quad j=1,2,\ldots n,\\
L=&&-2\frac{\partial^2}{\partial x^2}\left(\log\left(1+e^{{\eta_1+\eta_1^*+R}}\right)\right),
\eea
\label{Nwave}
where
\bea
\eta_1=k_1x-(ik_1^2+\omega_1)y+\omega_1t, \quad e^R=\frac{-\sum_{j=1}^{n}(\alpha_1^{(j)}\alpha_1^{(j)*})}{4k_{1R}\omega_{1R}}.
\eea
\endnumparts
\underline{(ii) Two soliton solution:}\\
Similar procedure results in the two soliton solution for the multicomponent case with arbitrary $(n+1)$ waves whose expression can be obtained from equations (14) by just allowing $j$ to run from $1,2,\ldots, n$ and redefining $\kappa_{il}$ as $\kappa_{il}=-\sum_{j=1}^{n}(\alpha_i^{(j)}\alpha_l^{(j)*})/(\omega_i+\omega_l^*)$, $i,l=1,2$. 
Now it is straightforward to extend the bilinearization procedure of obtaining one and two soliton solutions to obtain  multisoliton solutions as in the 1D integrable CNLS equations \cite{ref16}.  Similarly, three- and four soliton solutions of equation (26) can be obtained by suitably redefining $\kappa_{il}$'s with $i,l=1,2,3$ and $i,l=1,2,3,4$, respectively, and fixing the upper limit of the index $j$, corresponding to the short wave components as $n$.  The multisoliton solution of the  multicomponent case (26) can be written down from equation (6) by allowing $s$ to run from 1 to $n$ and redefining the column matrix $\psi_j$ as 
\bea
\psi_j=\left(\alpha_j^{(1)}, \alpha_j^{(2)}, \ldots, \alpha_j^{(n)} \right)^T.\nonumber
\eea
Likewise the proof of the multisoliton solution of the multicomponent system also follows the three wave system discussed in section 3.
\section{Conclusion}
To conclude, we have obtained explicitly the multi bright plane soliton solutions of recently reported physically interesting integrable (2+1) dimensional $(n+1)$-wave system by applying Hirota's bilinearization procedure.  We have also presented the results in a Gram determinant form for the multisoliton solutions of the multicomponent LSRI system along with the necessary proof.  We observe that the solitons in the short wave components can be amplified by merely reducing the pulse width of the long wave component.  The study on collision dynamics shows that the solitons appearing in the short wave components undergo shape changing collisions with intensity redistribution and amplitude-dependent phase shift.  This gives the exciting possibility of soliton collision based computing in higher dimensional integrable systems also.  However, the solitons in the long wave component always undergo elastic collision.
\ack
T. K. acknowledges the support of Department of Science and Technology, Government of India under the DST Fast Track Project for young scientists.  T. K. and K. S. thank the Principal and Management of Bishop Heber College, Tiruchirapalli, for constant support and encouragement.  The works of M. V. and M. L. are supported by a DST-IRPHA project.  M. L. is also supported by a DST Ramanna Fellowship. 

\noappendix

\section*{References}  
\begin{thebibliography}{20}
\bibitem{ref1}
Kivshar Y S and Agrawal G P 2003 {\it  Optical Solitons: From Fibers to Photonic Crystals} (San Diego: Academic Press)
\bibitem{ref2}
Manakov S V 1974 {\it Sov. Phys. JETP} {\bf 38} 248 
\bibitem{ref3}
Ablowitz M J, Prinari B and Trubatch T D 2004 {\it Discrete and Continuous Nonlinear Schr\"odinger Systems} (Cambridge: Cambridge University Press)
\bibitem{ref3a}
Radhakrishnan R, Lakshmanan M and Hietarinta J 1997 {\it Phys. Rev. E}
{\bf56} 2213
\bibitem{ref4}
Kanna T and Lakshmanan M 2001 {\it Phys. Rev. Lett.} {\bf 86} 5043
\bibitem{ref5}
Kanna T and Lakshmanan M 2003 {\it Phys. Rev. E} {\bf 67} 046617 
\bibitem{ref6}
Vijayajayanthi M, Kanna T and Lakshmanan M 2008 {\it Phys. Rev. A} {\bf 77} 013820 
\bibitem{ref16}
Kanna T, Lakshmanan M, Dinda P T and Akhmediev N 2006 {\it Phys. Rev. E} {\bf 73} 026604    
\bibitem{ref6a}  
Ablowitz M J, Ohta Y and Trubatch A D 1999 {\it Phys. Lett. A} {\bf 253} 287 
\bibitem{ref7}
Park Q H and Shin H J 2000 {\it Phys. Rev. E} {\bf 61} 3093 
\bibitem{ref8}
Hioe F T 2002 {\it J. Math. Phys.} {\bf 43} 6325 
\bibitem{ref10}
Sukhorukov A A and Akhmediev N N 2003 {\it Opt. Lett.} {\bf 28} 908 
\bibitem{ref10a}
Babarro J, Paz-Alonso M J, Michinel H, Salgueiro J R and Olivieri D N 2005 {\it  Phys. Rev. A} {\bf 71} 043608 
\bibitem{ref11}
Jakubowski M H, Steiglitz K and Squier R 1998 {\it Phys. Rev. E} {\bf
58} 6752    
\bibitem{ref12}
Steiglitz K 2000 {\it Phys. Rev. E} {\bf 63} 016608   
\bibitem{ref13}
Ohta Y, Maruno K and Oikawa M 2007 {\it J. Phys. A: Math. Theor.} {\bf 40} 7659  
\bibitem{ref13a}
Degasperis A, Conforti M, Baronio F and Wabnitz S 2006 {\it Phys. Rev. Lett.} {\bf 97} 093901 
\bibitem{ref14}
Oikawa M, Okamura M and Funakoshi M 1989 {\it J. Phys. Soc. Japan} {\bf 58} 4416   
\bibitem{radha}
Radha R, Kumar C S, Lakshmanan M, Tang X Y and Lou S Y 2005 {\it J. Phys. A: Math. Gen.} {\bf 38} 9649
\bibitem{ref15}
Hirota R 1973 {\it J. Math. Phys.} {\bf 14} 805 
\bibitem{hirotabook}
Hirota R 2004 {\it  The Direct Method in Soliton Theory} (Cambridge: Cambridge University Press)
\bibitem{proof}
Maruno K and Ohta Y 2008 {\it Phys. Lett. A} {\bf 372} 4446 
\end{thebibliography}
\end{document}